\documentclass[aps,prl,preprint,tightenlines,superscriptaddress,showpacs]{revtex4}

\usepackage{graphicx}
\usepackage{dcolumn}
\usepackage{bm}

\begin{document}

\vspace*{-3\baselineskip}
\resizebox{!}{3cm}{\includegraphics{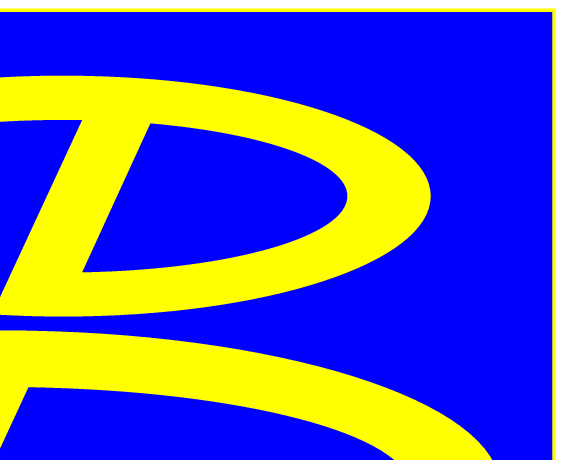}}
\preprint{BELLE Preprint 2002-19}

\preprint{KEK Preprint 2002-61}

\preprint{DPNU-02-20}

\title{Studies of the Decay $B^{\pm} \to D_{CP} K^{\pm}$}





\affiliation{Aomori University, Aomori}
\affiliation{Budker Institute of Nuclear Physics, Novosibirsk}
\affiliation{Chiba University, Chiba}
\affiliation{Chuo University, Tokyo}
\affiliation{University of Cincinnati, Cincinnati OH}
\affiliation{University of Frankfurt, Frankfurt}
\affiliation{Gyeongsang National University, Chinju}
\affiliation{University of Hawaii, Honolulu HI}
\affiliation{High Energy Accelerator Research Organization (KEK), Tsukuba}
\affiliation{Hiroshima Institute of Technology, Hiroshima}
\affiliation{Institute of High Energy Physics, Chinese Academy of Sciences, Beijing}
\affiliation{Institute of High Energy Physics, Vienna}
\affiliation{Institute for Theoretical and Experimental Physics, Moscow}
\affiliation{J. Stefan Institute, Ljubljana}
\affiliation{Kanagawa University, Yokohama}
\affiliation{Korea University, Seoul}
\affiliation{Kyoto University, Kyoto}
\affiliation{Kyungpook National University, Taegu}
\affiliation{Institut de Physique des Hautes \'Energies, Universit\'e de Lausanne, Lausanne}
\affiliation{University of Ljubljana, Ljubljana}
\affiliation{University of Maribor, Maribor}
\affiliation{University of Melbourne, Victoria}
\affiliation{Nagoya University, Nagoya}
\affiliation{Nara Women's University, Nara}
\affiliation{National Kaohsiung Normal University, Kaohsiung}
\affiliation{National Lien-Ho Institute of Technology, Miao Li}
\affiliation{National Taiwan University, Taipei}
\affiliation{H. Niewodniczanski Institute of Nuclear Physics, Krakow}
\affiliation{Nihon Dental College, Niigata}
\affiliation{Niigata University, Niigata}
\affiliation{Osaka City University, Osaka}
\affiliation{Osaka University, Osaka}
\affiliation{Panjab University, Chandigarh}
\affiliation{Peking University, Beijing}
\affiliation{Princeton University, Princeton NJ}
\affiliation{RIKEN BNL Research Center, Brookhaven NY}
\affiliation{Saga University, Saga}
\affiliation{University of Science and Technology of China, Hefei}
\affiliation{Seoul National University, Seoul}
\affiliation{Sungkyunkwan University, Suwon}
\affiliation{University of Sydney, Sydney NSW}
\affiliation{Tata Institute of Fundamental Research, Bombay}
\affiliation{Toho University, Funabashi}
\affiliation{Tohoku Gakuin University, Tagajo}
\affiliation{Tohoku University, Sendai}
\affiliation{University of Tokyo, Tokyo}
\affiliation{Tokyo Institute of Technology, Tokyo}
\affiliation{Tokyo Metropolitan University, Tokyo}
\affiliation{Tokyo University of Agriculture and Technology, Tokyo}
\affiliation{Toyama National College of Maritime Technology, Toyama}
\affiliation{University of Tsukuba, Tsukuba}
\affiliation{Utkal University, Bhubaneswer}
\affiliation{Virginia Polytechnic Institute and State University, Blacksburg VA}
\affiliation{Yokkaichi University, Yokkaichi}
\affiliation{Yonsei University, Seoul}
  \author{K.~Abe}\affiliation{High Energy Accelerator Research Organization (KEK), Tsukuba} 
  \author{K.~Abe}\affiliation{Tohoku Gakuin University, Tagajo} 
  \author{N.~Abe}\affiliation{Tokyo Institute of Technology, Tokyo} 
  \author{R.~Abe}\affiliation{Niigata University, Niigata} 
  \author{T.~Abe}\affiliation{Tohoku University, Sendai} 
  \author{Byoung~Sup~Ahn}\affiliation{Korea University, Seoul} 
  \author{H.~Aihara}\affiliation{University of Tokyo, Tokyo} 
  \author{M.~Akatsu}\affiliation{Nagoya University, Nagoya} 
  \author{Y.~Asano}\affiliation{University of Tsukuba, Tsukuba} 
  \author{T.~Aso}\affiliation{Toyama National College of Maritime Technology, Toyama} 
% ...
\collaboration{The Belle Collaboration}
\noaffiliation

\begin{abstract}
We report studies of the decay $B^{\pm} \to D_{CP} K^{\pm}$, 
where $D_{CP}$ denotes neutral $D$ mesons that decay to $CP$ eigenstates.
The analysis is based on a 29.1$~{\rm fb^{-1}}$ data sample of 
collected at the $\Upsilon(4S)$ resonance 
with the Belle detector at the KEKB asymmetric  $e^+ e^-$ storage ring. 
Ratios of branching fractions of Cabibbo-suppressed 
to Cabibbo-favored processes 
involving $D_{CP} $ are determined to be ${\cal B}(B^- \to D_1 K^-)/{\cal B}(B^- \to D_1 \pi^-)=0.125 \pm 0.036\pm0.010$ and ${\cal B}(B^- \to D_2 K^-)/{\cal B}(B^- \to D_2 \pi^-)=0.119\pm0.028\pm0.006$, 
where indices 1 and 2 represent the $CP=+1$ and $CP=-1$ eigenstates 
of the $D^0-\bar{D^0}$ system, respectively. 
We also extract the partial rate asymmetries 
for  $B^{\pm} \to D_{CP} K^{\pm}$, finding 
${\cal{A}}_1 = 0.29 \pm 0.26 \pm 0.05$ and  
${\cal{A}}_2 = -0.22 \pm 0.24 \pm 0.04$.
\end{abstract}

\pacs{12.15.Hh, 13.25.Hw}


\maketitle

\begin{figure}[!htb]
 \resizebox{0.8\textwidth}{!}{\includegraphics{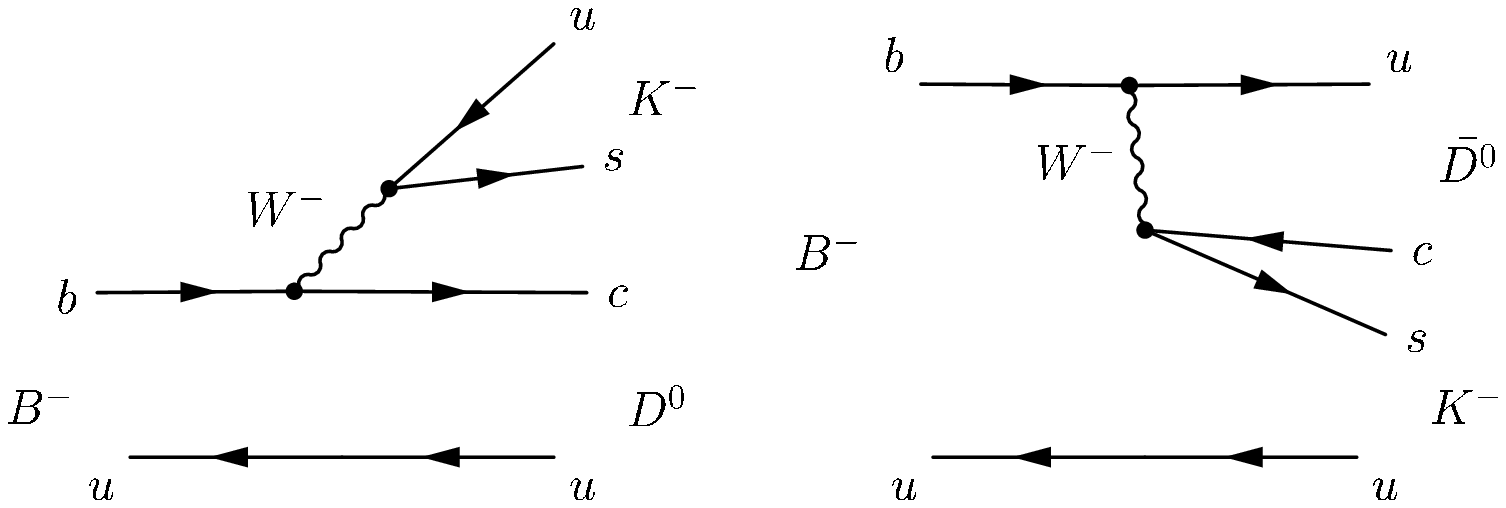}}%
 \caption{\label{dk}
 $B^- \to D^0 K^-$ decay amplitudes.
 }
\end{figure}

Direct $CP$-violation in $B^{\pm} \to D_{CP} K^{\pm}$ decay, 
where $D_{CP}$ denotes neutral $D$ mesons that decay to $CP$ eigenstates, 
provides a promising way to extract the angle $\phi_3$ 
of the Cabibbo-Kobayashi-Maskawa unitarity
triangle~\cite{gw,ads}. A partial rate asymmetry 
between the $D_{CP} K^-$ and $D_{CP} K^+$ final states 
can arise from interference between $b \to c$ and $b \to u$ 
processes shown in Fig.~\ref{dk}. 
The relation of the partial rate asymmetry ${\cal A}_{CP}$ and 
the $\phi_3$ angle~\cite{phi3} is given by: 
\begin{eqnarray}
 {\cal{A}}_{1,2} &\equiv& \frac{{\cal B}(B^- \to D_{1,2}K^-) - {\cal B}(B^+ \to D_{1,2}K^+) }
                {{\cal B}(B^- \to D_{1,2}K^-) + {\cal B}(B^+ \to D_{1,2}K^+) } \nonumber \\ 
&=&\frac{2 r \sin \delta ' \sin \phi_3}{1 + r^2 + 2 r \cos \delta ' \cos \phi_3}, \label{eq:Acp}
\end{eqnarray}
where indices 1 and 2 denote the $CP=+1$ and $CP=-1$ eigenstates of 
the neutral $D$ mesons, $r$ is the ratio of the amplitudes, 
$r \equiv | A(B^- \to {\bar D}^0 K^-)/A(B^- \to D^0 K^-) |$, 
and $\delta'$ is the strong phase difference between the two amplitudes, 
with $\delta' = \delta$ for $D_1$ and $\delta' = \delta + \pi$ for $D_2$. 
This asymmetry can have a non-zero value when both $\phi_3$ and $\delta$ 
are non-zero. In principle, one can constrain 
the angle $\phi_3$ from the measurement of asymmetries 
${\cal{A}}_{1}$ and ${\cal{A}}_{2}$.

Experimentally, $B \to DK$ processes have been studied using measurements 
of the ratio of the Cabibbo-suppressed process $B^- \to D^0 K^-$ to the 
Cabibbo-favored process $B^- \to D^0 \pi^-$. 
Belle~\cite{belledk} measured $ R = {\cal B}(B^- \to D^0 K^-)/{\cal B}(B^- \to D^0 \pi^-) =0.079 \pm 0.009 \pm 0.006$, 
while CLEO~\cite{cleo} reported $R =  0.055 \pm 0.014 \pm 0.005$. 
Assuming factorization, $R$ is naively expected to be 
$\tan^2\theta_C(f_K/f_\pi)^2 \approx 0.074$ 
in a tree-level approximation, where $\theta_C$ is the Cabibbo angle,
and $f_K$ and $f_\pi$ are the decay constants. Both measurements 
are in agreement with this theoretical expectation. 

In this Letter, we report the first measurement of  
$B^{\pm} \to D_{CP} K^{\pm}$ decay. We also give a 
comparison of the ratio of branching fractions $R$ 
for the flavor specific state and the $CP= \pm 1$ eigenstates, 
and a determination of the asymmetries ${\cal{A}}_{1,2}$. 
The results are base on a $29.1~{\rm fb^{-1}}$ data sample collected 
on the $\Upsilon(4S)$ resonance with the Belle detector~\cite{belle} 
at the KEKB asymmetric $e^+ e^-$ collider~\cite{kekb}. 
This corresponds to approximately 31.3 million $B \bar{B}$ events. 
The inclusion of charged conjugate states is implied 
throughout this Letter unless otherwise stated. 

Belle is a general-purpose detector with a 1.5 T superconducting 
solenoid magnet that can distinguish the Cabibbo-suppressed
process $B^- \to D^0 K^-$ from the Cabibbo-favored 
process $B^- \to D^0 \pi^-$ by means of particle identification 
and kinematic separation. A charged-particle tracking system, 
covering approximately $90\%$ of the solid angle 
in the center-of-mass~(cm) frame, is comprised of 
a three-layer silicon vertex detector~(SVD) and a 50-layer 
central drift chamber~(CDC). Identification of charged hadron species 
is accomplished by combining responses from silica aerogel 
\v{C}erenkov counters~(ACC), the time-of-flight detector~(TOF) and 
the $dE/dx$ measurement from the CDC; 
it provides more than $2.5\sigma$ $K/\pi$ separation 
for laboratory momenta up to $3.5~{\rm GeV/}c$. 
A CsI(Tl) Electromagnetic Calorimeter~(ECL) located 
inside the solenoid coil is used for photon~($\gamma$) detection. 
A detailed description of the Belle detector 
can be found elsewhere~\cite{belle}. 

We analyze both $B^- \to D^0 K^-$ and $B^- \to D^0 \pi^-$, 
which are collectively referred to as $B^- \to D^0 h^-$ decays. 
The processes $B^- \to D^0 K^-$ and $B^- \to D^0 \pi^-$ are distinguished 
by a tight requirement on the particle identification of the prompt $h^-$ 
($K^-$ or $\pi^-$) and the effect of the mass difference at the final 
stage of the event selection. 
The decay $B^- \to D^0 \pi^-$ is used
as a control sample to establish constraints on kinematic
variables, resolution of detectors, evaluation of systematic
uncertainties and normalization of results. 

Flavor specific $D^0$ meson candidates are reconstructed 
via $D_f \to K^- \pi^+$; for $CP=+1$ eigenstates we use 
$D_1 \to K^- K^+$ and $\pi^- \pi^+$ and for $CP=-1$ we use 
$D_2 \to K_S \pi^0$, $K_S \omega$, $K_S\phi$, $K_S\eta$ and $K_S\eta'$.

The $K_S \to \pi^+ \pi^-$ candidates are reconstructed from two 
oppositely charged tracks with an invariant mass 
within $\pm3\sigma$ of the nominal $K_S$ mass. 
Candidate $\pi^0$ mesons are reconstructed from pairs of $\gamma$'s, 
each with energy greater than $30~{\rm MeV}$, and 
are required to have a reconstructed mass within 
$\pm3\sigma$ of the nominal $\pi^0$ mass. 
We form candidate $\eta$ and $\eta'$ mesons using the 
$\gamma \gamma$ and $\eta \pi^+ \pi^-$ decay modes 
with mass ranges of $0.495~{\rm GeV/}c^2<M(\gamma \gamma)<0.578~{\rm GeV/}c^2$ 
and $0.904~{\rm GeV/}c^2<M(\eta \pi^+ \pi^-)<1.003~{\rm GeV/}c^2$, 
respectively. 
The $\eta$ momentum is required to be greater than $500~{\rm MeV/}c$. 

Candidate $\omega$ and $\phi$ mesons are reconstructed 
from $\pi^+\pi^-\pi^0$ and $K^+K^-$ combinations 
with invariant masses in the ranges  
$0.733~{\rm GeV/}c^2<M(\pi^+ \pi^- \pi^0)<0.819~{\rm GeV/}c^2$ and 
$1.007~{\rm GeV/}c^2<M(K^+K^-)<1.031~{\rm GeV/}c^2$, respectively. 
A helicity angle requirement for the $\omega$ and 
$\phi$ mesons of $|\cos\theta_{hel}|>0.4$ reduces 
the contributions from non-resonant 
$D^0 \to K_S \pi^+\pi^-\pi^0$ and $K_S K^+K^-$ decays to a negligible level.
For $D^0 \to K_S\omega$, the $D^0 \to K^{*-} \rho^+$ background 
is explicitly rejected by vetoing any $K_S \pi^-$ combination 
that forms an invariant mass 
within $\pm75~{\rm MeV/}c^2$ of the nominal $K^*$ mass. 

For each charged track from the $D^0$ meson decay, the particle identification 
system is used to determine a $K/\pi$ likelihood ratio 
$P(K/\pi) = {\cal L}_K/({\cal L}_K + {\cal L}_{\pi})$, 
where ${\cal L}_K$ and ${\cal L}_{\pi}$ are kaon and pion 
likelihoods~\cite{belle}. 
For $D^0 \to K^- \pi^+$ , $K^-K^+$, and $\pi^- \pi^+$, 
the charged tracks with $P(K/\pi)<0.7$ 
are assigned to be pions, while kaons are required to satisfy 
$P(K/\pi)>0.3$ for $D^0 \to K^- \pi^+$ and 
$P(K/\pi)>0.7$ for $D^0 \to K^-K^+$. 
Pions from candidate $D^0 \to \pi^- \pi^+$ decays are required to 
have momentum greater than $0.8~{\rm GeV/}c$, 
and the candidate is vetoed if either pion, 
when combined with any other track in the event, has an invariant mass 
that is within $\pm 50~{\rm MeV/}c^2$ of the nominal $J/\psi$ mass, 
or within $\pm 20~{\rm MeV/}c^2$ of the nominal $D^0$ mass. 

Candidate $D^0$ mesons are also required to have an 
invariant mass within $\pm2.5\sigma$ of the nominal $D^0$ mass, 
where $\sigma$ is the measured mass resolution, which ranges 
from $5$ to $18~{\rm MeV/}c^2$, depending on the decay channel. 
Tracks and photons from the $D^0$ candidate final states, 
except for $K_S \pi^0$, $K_S \eta$ and $K_S \eta'$, 
are refitted according to the nominal 
$D^0$ meson mass hypothesis and the reconstructed vertex position, 
in order to improve momentum determination. 

We reconstruct $B^- \to Dh^-$ events using 
the laboratory constrained mass~($M_{lc}$). $M_{lc}$ is the $B$ 
candidate mass calculated from the laboratory momenta by 
using an $e^+e^- \to B\bar{B}$ hypothesis: 
$M_{lc} = \sqrt{(E_B^{lab})^2 - (p_B)^2}$, where  
$p_B$ is the $B$ candidate's laboratory momentum vector 
and $E_B^{lab} =\frac{1}{E_{ee}}(s/2 + {\bf P_{{\it ee}} \cdot P_{\it B}})$, 
where $s$ is square of the cm energy, 
${\bf P_{\it B}}$ is the laboratory momentum vector of 
the $B$ meson candidate, 
and ${\bf P_{{\it ee}}}$ and $E_{ee}$ are the laboratory momentum  
and energy of the $e^+e^-$ system, respectively. 
$M_{lc}$ is independent of the mass assumption 
of the particle to be used, and is suitable for both 
$B^- \to D\pi^-$ and $B^- \to DK^-$ simultaneously.
We accept $B$ candidates with $5.27~{\rm GeV/}c^2<M_{lc}<5.29~{\rm GeV/}c^2$. 

Background events from  $e^+e^- \to q\bar{q}$ continuum processes 
are rejected using event shape variables that distinguish between 
spherical $B{\bar B}$ events and jet-like continuum events. 
We construct a Fisher discriminant, 
${\cal F} = \sum_{l=2,4} \alpha_l R_l^{so} + \sum_{l=1}^4 \beta_l R_l^{oo}$, 
where $\alpha_l$, $\beta_l$ are optimized coefficients to maximize
the discrimination between $B\bar{B}$ events and continuum events, and 
$R_l^{so}$, $R_l^{oo}$ are modified Fox-Wolfram moments~\cite{sfw}
\cite{sfw2}. Furthermore, the variable $\cos \theta_B$ is used, 
where $\theta_B$ is the angle between the $B$ flight direction in the 
$\Upsilon$(4S) rest frame and the beam axis. 
A single likelihood variable is formed from the 
probability density functions of ${\cal{F}}$ and $\cos \theta_B$.
The likelihood ratio is defined as 
${\cal LR} = {\cal L_{\rm{sig}}}/({\cal L_{\rm{sig}}} + {\cal L_{\rm{cont}}})$ 
where ${\cal L_{\rm{sig}}}$ and ${\cal L_{\rm{cont}}}$ are 
likelihoods defined by ${\cal{F}}$ and $\cos \theta_B$ for 
the signal and continuum background, respectively. 
Since each $D^0$ sub-decay mode has different backgrounds, 
we optimize the cut of the likelihood ratio for each mode by maximizing 
$S/\sqrt{S+N}$, where $S$ and $N$ denote the numbers of signal and 
background events estimated by a Monte Carlo simulation~\cite{mc}. 
For the $D^0 \to K^- \pi^+$ mode, the requirement ${\cal LR} > 0.4$ 
retains $87.1\%$ of signal and $26.4\%$ of the continuum background.

\begin{figure}[!htb]
\centerline{
 \resizebox{0.9\textwidth}{!}{\includegraphics{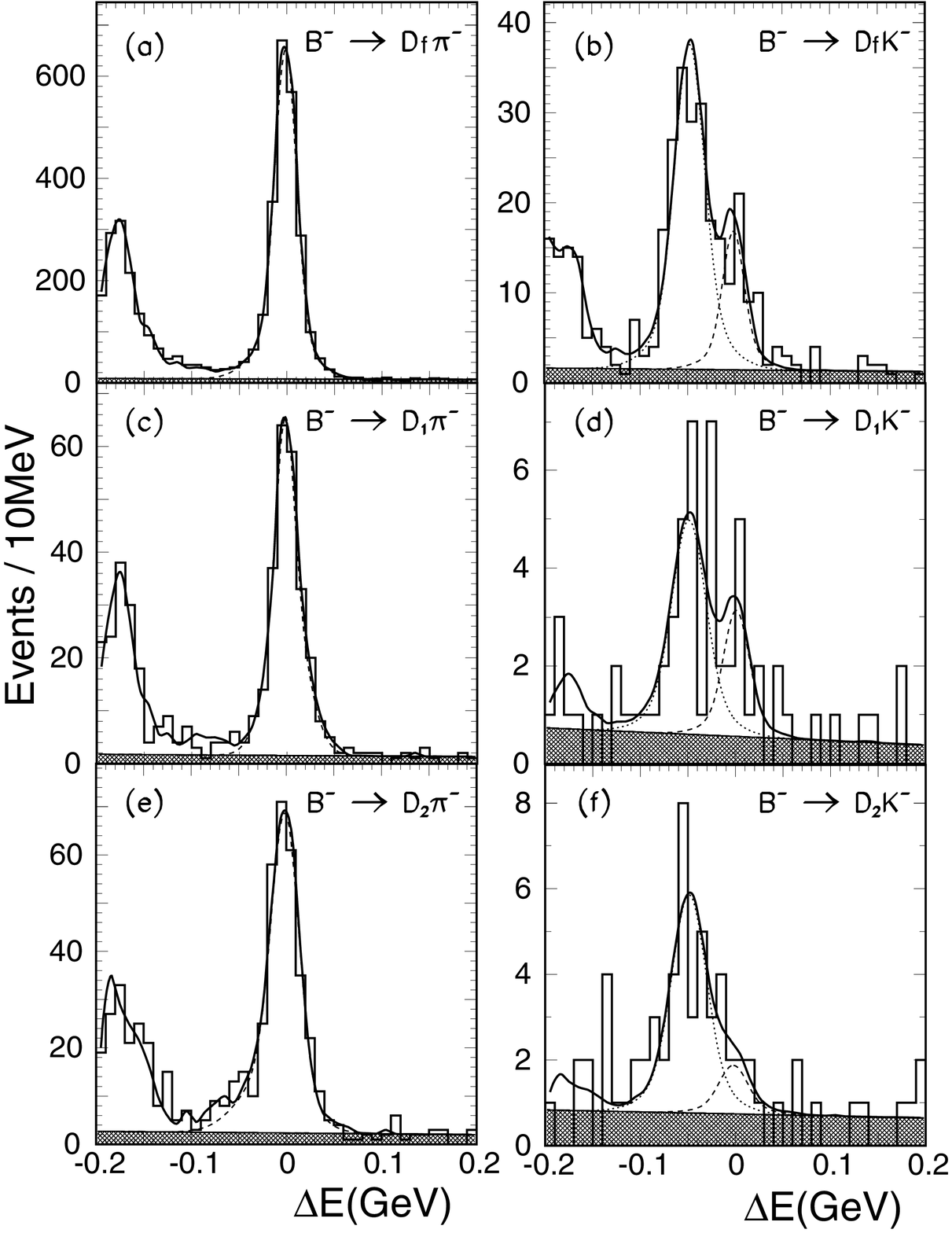}}%
}
\caption{ $\Delta E$ distributions for $B^- \to D^0 \pi^-/K^-$ 
candidates and fit results: (a) $B^- \to D_{f} \pi^-$, 
(b) $B^- \to D_{f} K^-$, (c) $B^- \to D_1 \pi^-$, (d) $B^- \to D_1 K^-$, 
(e) $B^- \to D_2 \pi^-$, and (f) $B^- \to D_2 K^- $, where 
in each case the pion mass is assigned to the prompt $\pi^-/K^-$. 
Dotted (dashed) lines show the distributions of $DK (D\pi)$ signals. 
The shaded plot shows the continuum background and the remaining 
component from $B\bar{B}$ background is estimated and fitted 
by Monte Carlo simulation. In the $DK$ plots, 
the dashed curves show the $D^0 \pi$ feed-across.}
\label{plotdpidk}
\end{figure}

To distinguish between $B^- \to D^0 K^-$ and $B^- \to D^0 \pi^-$ processes, 
we form two samples by requiring that the prompt hadron
$h^-$ satisfies either $P(K/\pi) > 0.8$ ($B^- \to D^0 K^-$ enriched sample) 
or $P(K/\pi) < 0.8$ ($B^- \to D^0 \pi^-$ enriched sample). 
Finally, the $\Delta E$ distribution, calculated 
by assigning the pion mass to the prompt hadron $h^-$, 
is examined to distinguish between $B^- \to D^0 K^-$ 
and $B^- \to D^0 \pi^-$ processes, 
where $\Delta E$ is the energy difference in the cm frame: 
$\Delta E = E_D^{cm} +E_{h^-}^{cm} - E_{{\rm beam}}^{cm}$. 
The signals for  $B^- \to D^0 \pi^-$ and $B^- \to D^0 K^-$ 
peak in the $\Delta E$ distributions at $0~{\rm MeV}$ 
and $-49~{\rm MeV}$ respectively, as can be seen in 
Fig.~\ref{plotdpidk}. In the $B^- \to D^0 K^-$ enriched sample, 
we can also see a peak around $\Delta E = 0~{\rm MeV}$ due to 
misidentified pions from $B^- \to D^0 \pi^-$ 
in Fig.~\ref{plotdpidk}(b), (d), and (f). 

The numbers of $B^- \to D^0 \pi^-$ and $B^- \to D^0 K^-$ events 
are extracted by fitting double Gaussian functions 
with different central values and widths to the $\Delta E$ distribution. 
Backgrounds originate from $q{\bar q}$ continuum events and $B\bar{B}$ events.
Continuum events are distributed over the entire $\Delta E$ region, 
and the shape of this background is determined by fitting 
a linear function to the $\Delta E$ distribution 
in the $M_{lc}$ sideband region
~($5.20~{\rm GeV/}c^2<M_{lc}<5.26~{\rm GeV/}c^2$). 
$B\bar{B}$ backgrounds, such as $B^- \to D^0 \rho^-$ and 
$B \to D^{*} \pi^-$ processes, are mostly seen at negative values of 
$\Delta E$; MC simulations are used to obtain 
the shape of their distribution. 

In the fits to the $B^- \to D^0 \pi^-$ enriched $\Delta E$ 
distribution, the signal peak position and width, 
and the normalization of continuum and 
$B\bar{B}$ backgrounds are free parameters. 
On the other hand, we calibrated the shape parameters of the 
$B^- \to D^0 K^-$ signal using the $B^- \to D^0 \pi^-$ data following 
the procedure described in \cite{belledk}, which accounts for the kinematical 
shifts and smearing of the $\Delta E$ peaks caused 
by the incorrect mass assignments of prompt hadrons. 
For the $B^- \to D^0 K^-$ fit, 
the shape parameters of the double Gaussian of the signal are fixed 
at values determined from the fit to the $B^- \to D^0 \pi^-$ sample. 
The peak position and width of the $B^- \to D^0 K^-$ signal events 
are determined by fitting the $B^- \to D^0 \pi^-$ distribution using a 
kaon mass hypothesis for the prompt pion, where the relative peak 
position is reversed with respect to the origin. 
The shape parameters for the feed-across from $B^- \to D^0 \pi^-$ is 
fixed by the fit results of the $B^- \to D^0 \pi^-$ enriched sample.

The fit results are shown as solid curves in Fig.~\ref{plotdpidk}.  
The number of events in the $D^0 K^-$ signal 
and $D^0 \pi^-$ feed-across, and the statistical significance 
of the $D^0 K^-$ signal are given in Table~\ref{event}. 
The statistical significance is defined as 
$\sqrt{-2\ln({\cal L}_0/{\cal L}_{max})}$, 
where ${\cal L}_{max}$ is the maximum likelihood 
in the $\Delta E$ fit and ${\cal L}_0$ is the likelihood when the signal 
yield is constrained to be zero. The statistical significance of 
both $D_1 K^-$ and  $D_2 K^-$ signals is over $5.0\sigma$. 

Experimentally, the ratio of branching fractions is determined as follows: 
\begin{eqnarray}
 R = \frac{N(B^- \to D^0 K^-)}{N(B^- \to D^0 \pi^-)} \times \frac{\eta(B^- \to D^0 \pi^-)}{\eta(B^- \to D^0 K^-)} \times \frac{\epsilon(\pi)}{\epsilon(K)},
\end{eqnarray}
where $N$ is the number of events obtained, $\eta$ is the signal 
detection efficiency, and $\epsilon$ is 
the prompt pion or kaon identification efficiency. 
The signal detection efficiencies are determined from 
MC simulation: $\eta(B^- \to D^0 K^-)$ is approximately $5\%$ lower than 
$\eta(B^- \to D^0 \pi^-)$ due to kaon decays in flight. 
Kaon and pion identification efficiencies 
$\epsilon (K)$ and $\epsilon (\pi)$ are experimentally determined 
from a kinematically selected sample of 
$D^{*+} \to D^0 \pi^+$ and $D^0 \to K^- \pi^+$ decays with $K^-$ and 
$\pi^+$ mesons from $D^0$ candidates that are in the same cm 
momentum~($p^{cm}$) and polar angle regions 
as prompt hadrons in the $B^- \to Dh^-$ decay 
($2.1~{\rm GeV/}c < p^{cm} <2.5~{\rm GeV/}c $). 
With our requirements, $P(K/\pi) > 0.8$ gives a kaon identification 
efficiency of $\epsilon (K)=0.778\pm0.005$ with a pion misidentification rate 
of $0.024\pm0.002$, and $P(K/\pi) < 0.8$ gives a pion 
identification efficiency of $\epsilon(\pi)=0.972\pm0.007$. 

Since the kinematics of the $B^- \to D^0 K^-$ and $B^- \to D^0 \pi^-$ 
processes are similar, the systematic uncertainties 
from detection efficiencies cancel in the ratios of 
branching fractions. The dominant systematic errors are the uncertainties in 
the shapes of backgrounds in the $\Delta E$ distributions~($5.1\% - 7.9\%$), 
which are determined by varying the background shape of the fitting 
function by $\pm 1\sigma$, and $K/\pi$ identification efficiencies~($1.2\%$). 
The sum of the uncertainties for each mode are combined in quadrature 
to determine the total systematic errors for the ratios.

The resulting measurements of $R$ are listed with their statistical 
and systematic errors in Table~\ref{event}. These are the first 
observations of the decays $B^- \to D_{CP}K^-$. As a check, 
the result for the flavor specific decay $B^- \to D^0 h^-, D^0 \to K^- \pi^+$ 
is listed as well, and is found to be consistent with 
previous measurements. We find no deviation of the $R$ ratio for 
the $B^- \to D_{CP} K^-$ processes from the corresponding flavor specific 
modes beyond statistical errors. 

Direct $CP$ violation is investigated by measuring the partial rate 
asymmetries in $B^{\pm} \to D_{1,2}K^{\pm}$ decays. 
We fit the yields of $B^+$ and $B^-$ events separately for 
each mode and determine the partial rate asymmetries 
${\cal A}_{1,2}$ are shown in Table~\ref{Atable}. 
The partial rate asymmetry is consistent with zero 
for the flavor specific mode, which is expected to have no asymmetry. 
To construct $90\%$ confidence intervals of 
${\cal A}_{1}$ and ${\cal A}_{2}$, we combine statistical 
and systematic errors in quadrature, and assume that the total
error is distributed as a Gaussian.  
We find $-0.14<{\cal A}_1<0.73$ and $-0.62<{\cal A}_2<0.18$, 
consistent with zero asymmetry.

The main sources of systematic errors for the partial rate asymmetries 
${\cal A}_{1,2}$ are asymmetries in the measured 
background~($1.5\% - 3.9\%$), 
intrinsic asymmetry in the Belle detector~($3.6\%$), 
and kaon identification~($1.0\%$). We observe 
$1154.6\pm35.4$ $B^+ \to \bar{D}^0 \pi^+,\bar{D}^0 \to K^+ \pi^- $ 
candidates and $1073.5\pm34.5$ $B^- \to D^0 \pi^-, D^0 \to K^- \pi^+$ 
candidates. This is consistent with our expectation that 
the detector has no significant intrinsic charge asymmetry. 
Using MC simulation, we find that 
the contribution of the non-resonant contaminations~($\sim 0.1\%$) 
of $\omega$ and $\phi$ can be neglected. 

In conclusion, using $29.1~{\rm fb^{-1}}$ of data collected with the Belle 
detector, we report studies of the decays 
$B^{\pm} \to D_{CP} K^{\pm}$, where $D_{CP}$ are the neutral $D$ meson 
$CP$ eigenstates. This is the first stage of a program to 
measure the angle $\phi_3$ in the CKM unitarity triangle. 
The ratios of branching fractions $R$ 
for the decays $B^- \to D_{CP} K^-$ and $B^- \to D_{CP} \pi^-$ 
are consistent with that for the flavor specific decay within errors. 
The measured partial rate asymmetries ${\cal A}_{1,2}$ are 
consistent with zero within large errors. 
In the future, with $300~{\rm fb^{-1}}$ of data, 
an accuracy of better than 0.1 in the measurements of 
$R_{1,2}$ and ${\cal{A}}_{1,2}$ in $B^-\to D_{CP}K^-$ is
expected and we will begin to constrain the angle $\phi_3$. 


We wish to thank the KEKB accelerator group for the excellent
operation of the KEKB accelerator. We acknowledge support from the
Ministry of Education, Culture, Sports, Science, and Technology of Japan
and the Japan Society for the Promotion of Science; the Australian
Research
Council and the Australian Department of Industry, Science and
Resources; the Department of Science and Technology of India; the BK21
program of the Ministry of Education of Korea and the CHEP SRC
program of the Korea Science and Engineering Foundation; the Polish
State Committee for Scientific Research under contract No.2P03B 17017;
the Ministry of Science and Technology of Russian Federation; the
National Science Council and the Ministry of Education of Taiwan; the
Japan-Taiwan Cooperative Program of the Interchange Association; 
National Science Foundation of China under Contract No. 10175071; 
and the U.S. Department of Energy.

%
%
%

%
%

\begin{table*}
\caption{Results of fits for the $D^0 \pi^-$ and $D^0 K^-$ decay modes.  
The event yields, the feed-across from $D^0 \pi^-$ to 
the $D^0 K^-$ signal region, 
statistical significance of $D^0 K^-$ signals, efficiencies and 
branching fraction ratios($R$) are given. 
Efficiencies are determined by weighting according 
to the measured sub-components
($\eta \equiv \sum_{i} \eta_i {\cal B}(D^0 \to X_i)$).}
 \begin{ruledtabular}
  \catcode`;=\active \def;{\phantom{}}
\begin{center}
\begin{tabular}{lcccccc}
 & $B^- \to D^0 K^- $ & $B^- \to D^0 \pi^-$ & Stat. & $B^- \to D^0 \pi^-$  & Efficiency($\%$) & branching fraction ratio \\
 &  events              & feed-across           & sig. 	&  events & $\eta(D^0 \pi^-) / \eta(D^0 K^-)$ & $R$ \\
\hline
$ B^- \to D_{f} h^-$ & $161.7 \pm 14.5$ & $51.3 \pm 9.7$ & 16.9 & $2245.1 \pm 51.0$ & 1.703/1.639 & $0.094 \pm 0.009 \pm 0.007$ \\
$ B^- \to D_1 h^-$ & $22.9 \pm 6.1$ & $9.6 \pm 4.4$ & 5.1 & $240.1 \pm 16.7$ & 0.173/0.165 & $0.125 \pm 0.036 \pm 0.010$ \\
$ B^- \to D_2 h^-$ & $26.1 \pm 6.5$ & $4.9 \pm 4.1$ & 5.5 & $290.6 \pm 19.1$ & 0.184/0.173 & $0.119 \pm 0.028 \pm 0.006$ 
\end{tabular}
\end{center}
 \end{ruledtabular}
\label{event}
\end{table*}

\begin{table*}
\caption{Summary of measured partial rate asymmetries. }
 \begin{ruledtabular}
  \catcode`;=\active \def;{\phantom{}}
\begin{center}
\begin{tabular}{ccccc}
Mode   & $N(B^+)$ & $N(B^-)$ & ${\cal A}_{CP}$ & 90\% C.L.\\ 
\hline
$B^{\pm} \to D_{f} K^{\pm}$ & $80.6 \pm 10.1$ & $81.1 \pm 10.4$ & $0.003 \pm 0.089 \pm 0.037$ & $-0.15<{\cal A}_{f}<0.16$ \\
$B^{\pm} \to D_1 K^{\pm}$ & $8.1 \pm 3.9$ & $14.7 \pm 4.6$ & $0.29 \pm 0.26 \pm 0.05$ & $-0.14<{\cal A}_1<0.73$ \\
$B^{\pm} \to D_2 K^{\pm}$ & $16.4 \pm 5.5$ & $10.6 \pm 4.2$ & $-0.22 \pm 0.24 \pm 0.04$ & $-0.62<{\cal A}_2<0.18$ \\
\end{tabular} 
\end{center}
 \end{ruledtabular}
\label{Atable}
\end{table*}


\begin{thebibliography}{99}
\bibitem{df} The charge conjugated mode($\bar{D^0} \to K^+ \pi^-$) 
	is implicitly included.
\bibitem{helicity} The $\omega$ helicity angle is defined as 
	the angle between the $D^0$ flight direction in the $\Upsilon$(4S)'s 
	frame and the normal to the $\omega$ decay plane 
	in the $\omega$'s rest frame. 
	The $\phi$ helicity angle is defined as the angle 
	between the $\phi$ flight direction in the $D$'s rest frame 
	and the direction of the $\phi$'s daughter in the $\phi$'s rest frame.
\bibitem{sfw2} The index $so$($oo$) indicates that the moment is 
calculated from pairs of particles where only one (neither) particle 
comes from the $B$ candidate.
\end{thebibliography}
\end{document}